# Spectroscopy of the Double Quasars Q1343+266AB: A New Determination of the Size of Lyα Forest Absorbers

Jill Bechtold[1], Arlin P.S. Crotts[2], Robert C. Duncan[3], Yihu Fang[2]

## Abstract

We have obtained spectroscopy of Q 1343+266 AB, a pair of quasars at redshift $z = 2.03$ with a projected separation of 9.5 arcseconds. This system is well-suited for probing the Lyα forest, since the two component spectra show several Lyα lines in common and several others not. Using Bayesian statistics, under the idealization of uniform-radius spherical absorbers, we find that the Lyα cloud radius at $z \approx 1.8$ lies in the range $40\,h^{-1}$ kpc $< R < 280\,h^{-1}$ kpc with 98% confidence in a $(\Omega_o, \Lambda_o) = (1,0)$ model universe, where $h \equiv (H_o/100$ km s$^{-1}$ Mpc$^{-1})$. The median value of $R$ is $90\,h^{-1}$ kpc. These numbers scale up by a factor 1.44 for $(\Omega_o, \Lambda_o) = (0.1, 0)$; and by a factor 1.85 for $(0.1, 0.9)$. Pressure-confined and freely-expanding Lyα cloud models in which the absorbers formed in a significantly more compressed state at $z \gg 2$ are contradicted by these new data, as are models involving stably-confined gas concentrations in photoionization equilibrium within minihalos of cold dark matter. The comoving density of Lyα forest objects at $z \sim 2$ is $\sim 0.3\,h\,(R/100$ kpc$)^{-2}$ Mpc$^{-3}$ in a (1,0) model universe. This suggests a possible identification of Lyα clouds with the unvirialized, collapsing progenitors of the faint, blue galaxies at $z \sim 1$.

*Subject headings:* galaxies: quasars: aborption lines — galaxies: quasars: individual: Q 1343+266

## 1. Introduction

Spectral observations of gravitationally lensed quasars, or quasar pairs with small projected separations, are a key to understanding the Lyα forest (Oort 1981). By measuring how many Lyα lines are common in the two component spectra, and how equivalent widths of these common lines compare, one can constrain the transverse size and structure of the absorbers. Studies of quasar pairs with large separations (∼1 arcmin or larger) have detected very few common absorption lines, placing upper limits on the characteristic Lyα absorber size of $\sim 0.5$–$2h^{-1}$ Mpc (Sargent, Young & Schneider 1982, Shaver & Robertson 1983, Crotts 1989). Gravitational lens systems have been used to probe much smaller scales (Weymann & Foltz 1983, Smette et al. 1992). These authors found that virtually all of the verifiable Ly$\alpha$ lines in one image spectra are present in the other, and that the equivalent widths of these common lines are statistically equal, indicating that the cloud size is much larger than the transverse linear separation of 0–1 $h^{-1}$ kpc.

---

[1]Steward Observatory, University of Arizona, Tucson, AZ 85721

[2]Dept. of Astronomy, Columbia University, 538 W. 120th Street, New York, NY 10027

[3]Dept. of Astronomy, University of Texas, Austin, TX 78712.

Pairs of quasars or lenses with separations in the intermediate range of $10''$–$1'$ are thus expected to probe the actual sizes of the clouds. Q2345+007 has 2 images separated by $7.3''$, and was used by Foltz et al. (1984) to put an approximate lower bound of $\sim$ 5–25 kpc on the characteristic size of the Ly$\alpha$ absorbers with $\langle z \rangle$=1.95. Subsequent analyses of these data (McGill 1990; Bajtlik & Duncan 1991; Smette et al. 1992), have ultimately not placed reliable, two-sided bounds on the absorber sizes, because of large uncertainties about the lens redshift (Steidel & Sargent 1991, "SS"; Fischer et al. 1994) and because several new metal lines were discovered by SS in the Ly$\alpha$ forest after all of the above-quoted analyses were done, leaving only a single candidate Ly$\alpha$ line that is confidently *unmatched* between the two component spectra. Nevertheless, the Q2345+007 data give reliable lower bounds $R > 2\,h^{-1}$ kpc (at $\sim$ 97% confidence across the possible lens redshift range) to the absorber sizes.

In this paper we present new spectroscopy of the pair of quasars 1343+266 AB. These quasars are at nearly identical redshifts, $z = 2.03$, and a separation of $9.5''$. They were discovered in the Crampton-Cowley quasar survey at the CFHT (Crampton et al. 1988a). Spectroscopy of both quasars and a fruitless search for a lensing galaxy or cluster were presented by Crampton et al. 1988b. In an accompanying paper (Crotts, Bechtold, Fang & Duncan 1994, hereafter Paper II) we give additional evidence that Q1343+266 AB are distinct quasars, not lensed images of a single object, and discuss the metal-line systems. In this paper we present our observations and briefly outline implications for the Ly$\alpha$ clouds, which will be discussed further in another paper (Duncan, Fang, Crotts & Bechtold 1994, hereafter Paper III).

## 2. Observations and Ly$\alpha$ absorber size analysis

We observed the two quasars 1343+266AB on the nights of April 4-5, 1994, with the MMT Blue Spectrograph, Loral 3048 × 1028 CCD and the 800 l/mm grating used in first order, with a $1'' \times 180''$ slit. The slit was aligned with the parallactic angle for most observations, except near transit, when it was rotated to include both quasars at once. One-hour exposures were made, interspersed with calibration HeNeArCu lamp exposures. The dome was used for flat fields. Because of CCD traps and bad columns, the quasars were moved along the slit between each exposure by several arcseconds. Also, the grating tilt was changed between nights, moving the spectrum by about 100 pixels in dispersion. The summed HeNeArCu exposures were used to estimate the wavelengths resolution, with 0.75 Å per pixel, 2.5 pixels FWHM from 3200 to 4600 Å, corresponding to a resolution of $\sim$ 175 to 125 km s$^{-1}$ over this range. Figure 1 shows the spectra, with absorption lines indicated. Details of the analysis, and the line list, are given in Paper II.

The proper separation of raypaths of the two binary components Q1343+266 A and B, in the observed Ly$\alpha$ forest, is $S = 39\,h^{-1}$ kpc, $56\,h^{-1}$ kpc, or $72\,h^{-1}$ kpc for the three fiducial cosmological models, $(\Omega_o, \Lambda_o) = (1,0)$, $(0.1, 0)$ or $(0.1, 0.9)$. These values of $S$ change by only a few percent over the observed redshift range of the forest, since the quasars are a pair, not lensed (Paper II). To find statistical bounds on Ly$\alpha$ absorber sizes, we count the numbers of "hits" and "misses," $\mathcal{N}_h$ and $\mathcal{N}_m$, in the spectra. A "hit" occurs when Ly$\alpha$ lines are detected with $> 3.5\,\sigma$ confidence in both spectra at the same redshift, operationally defined by $\Delta V < 200$ km s$^{-1}$. (Given the $z$–depth and line density of our spectra, the expected number of *random* Ly$\alpha$ associations at $\Delta V < 200$ km s$^{-1}$ is $< 0.1$, so random hits can be neglected.) A "miss" occurs whenever two criteria are met: (1) a Ly$\alpha$ line is seen in one spectrum with $> 3.5\,\sigma$ confidence; and (2) an equal-strength line at the same redshift ($\pm 200$ km s$^{-1}$) is *excluded* in the other spectrum with $> 3.5\,\sigma$ confidence. By these criteria, we find $\mathcal{N}_h = 11$ and $\mathcal{N}_m = 4$. Several lines in the Ly$\alpha$ forest region are not used because they are identified with metal-line systems (in A: #2, 4, 14, 21, 23; in B: #2, 14; Paper II), and 3 line pairs are ambiguous and are not counted in our Ly$\alpha$ absorber size analysis (lines # 1, 3 and 5 of A, Paper II). We have included as "hits" 4 cases where there is possible contamination by a metal line system, but where we can argue from doublet ratios or other line ratios that the equivalent width is dominated by a metal-free Ly$\alpha$ system. Extrapolating from the spectrum redward of Ly$\alpha$ emission, we estimate that there is probably less than one unidentified metal-line (pair) in our Ly$\alpha$ line-list. We begin by including lines near the quasars, even though they may be influenced by the "proximity effect" (c.f. Bechtold 1994; Bajtlik et al. 1989), and also lines in the spectrum of A which are at the position of Ly$\alpha$ corresponding to the BAL CIV trough (Paper II). The effect of excluding these lines is discussed below (Table 1).

We now calculate bounds on the Ly$\alpha$ absorber sizes, using Bayes' theorem (e.g., Press 1989). Adopting the idealization of spherical clouds of uniform radius $R$, the *probability density for $R$*, given $\mathcal{N}_h$ hits and $\mathcal{N}_m$ misses over a raypath separation $S$, is (Paper III):

$$\mathcal{P}(R) = \frac{4}{\pi} \left( \frac{(\mathcal{N}_h + \mathcal{N}_m + 1)!}{\mathcal{N}_h!\ \mathcal{N}_m!} \right) \frac{X}{R} (1 - X^2)^{1/2}\ \phi^{\mathcal{N}_h}\ (1 - \phi)^{\mathcal{N}_m}, \qquad (1)$$

where $X \equiv S/2R$. In this equation, $\phi$ is the probability for a "hit" when at least one random raypath intersects the cloud (McGill 1990):

$$\phi = (2/\pi)\,[\cos^{-1} X - X(1 - X^2)^{1/2}] \qquad \text{for}\ \ X < 1, \qquad (2)$$

and $\phi = 0$ otherwise. In deriving eq. (1) we adopted the Bayesian prior distribution that all values of $R$ are equally likely, since all previous lensed/binary quasar observations have probed much larger or much smaller spatial scales, giving only extreme upper and lower bounds to the absorber sizes. If these bounds are included in the prior distribution, they do not significantly affect the results.

Figure 2 is a plot of $\mathcal{P}(R)$ [solid line] and its integral [dashed line], for $(\Omega_o, \Lambda_o) = (1, 0)$. The median value of $R$, for which there is equal probability that the true $R$ lies above or below, is $87\,h^{-1}$ kpc [peak of the dashed curve], while the single most probable value of $R$ is $70\,h^{-1}$ kpc [peak of solid curve]. The 99% confidence lower and upper bounds are $43\,h^{-1}$ kpc $< R < 270\,h^{-1}$ kpc. For other cosmological models, these numbers scale with $S$ [e.g., larger by 1.44 for $(0.1, 0)$; larger by 1.85 for $(0.1, 0.9)$].

We also carried out the same analysis on more conservative line lists, using only lines detected with $4\sigma, 4.5\sigma$ and $5\sigma$ confidence. In addition, we tried a sample where we excluded all lines redward of 3550 Å; this excludes any Ly$\alpha$ line which may be associated with the CIV BAL outflow (Paper II) or the proximity effect. Table 1 shows that the statistical bounds on cloud sizes are not very sensitive to these changes.

Comparing common lines in the spectra of A and B gives additional information about the spatial scale of Ly$\alpha$ absorbers. Here any possible partial contamination by metal line systems must be considered with care. The difference $|W_A - W_B|$ exceeds its measurement uncertainty for several lines (Paper II). There is no evidence in our data however for significant velocity shifts, $\Delta V$, between matched lines, although the uncertainties in the shifts are substantial, $\sigma_{\Delta V} \sim 30$ km s$^{-1}$ (see Table 1 of Paper II for precise numbers). The distribution of $\Delta V/\sigma_{\Delta V}$ is consistent with a Gaussian of dispersion unity, with a KS probability Q=0.175 that the intrinsic velocity shifts are zero.

## 3. Implications for the Ly$\alpha$ Forest

What do these large sizes mean for models of the Ly$\alpha$ forest clouds? One set of models involves clouds that are confined by a hot, intergalactic medium (Sargent et al. 1980; Ikeuchi & Ostriker 1986 "IO") or are freely expanding (Bond, Szalay & Silk 1988 "BSS"). Several versions of such models have been proposed. One classic version postulates that the Ly$\alpha$ clouds formed in a significantly more compressed state at $z \gg 2$, where the initial compression was effected by shocks (e.g., IO §3; Vishniac & Bust 1987; Madau & Meiksin 1991) or by gravitational collapse before the onset of photoionization (BSS). In such a scenario, the dimensionless parameter $\Psi = R/c_s\tau_H$ satisfies $\Psi \leq 1$, where $c_s = (kT/\mu)^{1/2}$ is the (isothermal) sound speed in primordial-composition, photoionized gas ($T \sim 10^4$ K) and $\tau_H$ is the Hubble time at the epoch of observation, $z \approx 2$. Adopting the "conservative" (in the sense of giving the models the most chance to succeed) value of temperature $T = 3 \times 10^4$ K, we find that $\Psi \leq 1$ can be ruled out in the three fiducial cosmological models with $> 99.9\%$, $99.8\%$ and $97.5\%$ confidence, respectively. Thus, the clouds could not have expanded fast enough to reach the observed size at $z \sim 2$, even if they expand unimpeded by any confining pressure. If, on the other hand, the clouds somehow *formed* with the present large sizes in pressure equilibrium with a general intercloud medium at $z \gg 2$, then they could not have expanded quickly enough to remain in pressure equilibrium at the epoch of observation (IO; Paper III).

Another set of models involves photoionized gas stably confined by "minihalos" of cold, dark matter (Rees 1986; Miralda-Escudé & Rees 1993, "MR," and references therein). These models, in their simplest forms, predict that the impact parameter (radius $R$) at which a raypath through a cloud with an isothermal density profile intercepts an HI column density $N_{14} \times 10^{14}$ cm$^{-2}$ is (MR; Paper III):

$$R\,[\text{minihalo}] = 20\,\text{kpc}\,\left(\frac{N_{14}}{0.8}\right)^{-1/3}\left(\frac{T}{3\times 10^4\,\text{K}}\right)^{5/2}\left(\frac{f_g}{0.05}\right)^{2/3}\left(\frac{J_{21}}{0.3}\right)^{-1/3}. \qquad (3)$$

where $f_g$ is the ratio of baryon gas to total mass in the Universe, and the metagalactic radiation field at the Lyman limit is $J_\nu = J_{21} \times 10^{21}$ ergs sec$^{-1}$ cm$^{-2}$ Hz$^{-1}$. Again, we have adopted conservative parameter values; in particular $N_{14} = 0.8$ corresponds to the *smallest* rest-frame equivalent width line in our sample ($W = 0.22$ Å) taken with the maximum plausible thermal velocity width ($b = 30$ km s$^{-1}$). Nevertheless, the value of $R$ in eq. (3) is about a factor 4 times smaller than our 99% confidence lower bound, $R > 86\,(h/0.5)^{-1}$ kpc, which argues against the standard CDM minihalo model.

If these models are excluded, then what might the Ly$\alpha$ forest absorbers be? The comoving density of Ly$\alpha$ forest absorbers at $z \approx 2$ can be estimated using the new transverse size bounds and the observed line number density per unit redshift, with essentially no model-specific assumptions. In an $(\Omega, \Lambda) = (1, 0)$ model universe this yields (Paper III):

$$n_{L\alpha} \approx 0.30 \ h^3 \ (R/100 \ h^{-1} \ {\rm kpc})^{-2} \ {\rm Mpc}^{-3}, \qquad (4)$$

for counting threshold $W > 0.3$ Å [ or larger by 1.7 for $W > 0.2$ Å]. This comoving density exceeds by $\sim 30$ the density of $L_*$ galaxies in the present epoch ($z = 0$), but it could be comparable to the comoving density of star-bursting dwarf galaxies at $z \sim 1$, if such objects are responsible for the faint blue galaxies (see Paper III for details and references). Since the timescale for dynamical collapse of overdensites of this scale at $z = 2$ is also comparable to the cosmic time difference between $z \sim 2$ and $z \sim 1$ (Paper III), we suggest that *the Ly$\alpha$ clouds at $z \sim 2$ are the dynamically collapsing progenitors of faint blue galaxies*. Note that the Ly$\alpha$ clouds and the blue galaxies at magnitudes $> 26$ show similar, weak clustering (Efstathiou et al. 1991; Ostriker et al. 1987).

In conclusion, we have presented new observations of the quasar pair Q1343+266AB and shown that the characteristic size of the Ly$\alpha$ forest clouds is large: $R \sim 90 \ h^{-1}$ kpc at $z \approx 1.8$. We suggest therefore that some widely discussed models for the clouds, which propose that they are dynamically-stable and persistently-confined, need to be re-examined. The pronounced redshift evolution of Ly$\alpha$ line numbers for $0 < z < 5$ may also be a sign of ongoing dynamical evolution.

It is a pleasure to thank the staff of the Multiple Mirror Telescope for their assistance in obtaining these observations. We thank C. Foltz, S. Bajtlik, M. Giroux, J. Ostriker, P. Shapiro, E. Turner and E. Vishniac for enlightening discussions. This work was supported by NSF grant AST-9020757 (to RD), AST-9116390 and the David and Lucile Packard Foundation (to AC) and AST-9058510 and a gift from Sun Microsystems (to JB).

## References

Bajtlik, S., Duncan, R.C. & Ostriker J.P. 1988, Ap.J. 327, 570.
Bajtlik, S. & Duncan, R.C. 1991, in Proceedings of the ESO Mini-Workshop on Quasar Absorption Lines, ed. P.A. Shaver et al. (ESO: Garching) p. 35.
Bechtold, J. 1994, ApJS, 91, 1.
Bond, J.R., Szalay, A.S. & Silk, J. 1988, ApJ, 324, 627 (BSS)
Crampton, D., Cowley, A.P., Hickson, P., Kindl, E., Wagner, R.M., Tyson, J.A., Gullixson, C., 1988b, Ap.J., 330, 184.
Crampton, D., Cowley, A.P., Schmidtke, P.C., Janson, T. & Durrell, P., 1988a, AJ, 96, 816.
Crotts, A., 1989, Ap.J., 336, 550.
Crotts, A.P.S., Bechtold, J., Fang, Y. & Duncan, R.C. 1994, ApJ, submitted (Paper II).
Duncan, R.C., Fang, Y., Crotts, A.P.S. & Bechtold, J. 1994, to be submitted to Nature (Paper III).
Efstathiou, G., Bernstein, G., Katz, N., Tyson, J.A. & Guhathakurta, P. 1991, Ap.J. 380, L47.
Fischer, P., Tyson, J.A., Bernstein, G.M., & Guhathakurta, P. 1994, Ap.J. 431, L71.
Foltz, C.B., Weymann, R.J., Roser, H-J., Chaffee, F.H., 1984, Ap.J. 281, L1.
Ikeuchi, S. & Ostriker, J.P. 1986, ApJ, 301, 522 (IO).
McGill, C. 1990, MNRAS, 242, 544.
Madau, P. & Meiksin, A. 1991, Ap.J. 374, 6

Miralda-Escudé, J., & Rees, M.J. 1993, MNRAS, 260, 617 (MR).
Oort, J.H., 1981, A.A., 94, 359.
Ostriker, J.P., Bajtlik, S. & Duncan, R.C. 1988, Ap.J. 327, L35.
Press, S.J. 1989 Baysian Statistics: Principles, Models & Applications (Wiley: N.Y.)
Rees, M.J. 1986, MNRAS, 218, 25p.
Sargent, W.L.W., Young, P., Boksenberg, A. & Tytler, D., 1980, ApJS, 42, 41.
Sargent, W.L.W., Young, P. & Schneider, D.P., 1982, Ap.J., 256, 374.
Shaver, P.A. & Robertson, J.G., 1983, Ap.J., 268, L57.
Smette, A., Surdej, J., Shaver, P.A., Foltz, C.B., Chaffee, F.H., Weymann, R.J., Williams, R.E., & Magain, P., 1992, Ap.J., 389, 39.
Steidel, C.C. & Sargent, W.L.W., 1991, A.J., 102, 1610.
Vishniac, E.T. & Bust, G.S. 1987, Ap.J. 319, 14.
Weymann, R.J. & Foltz, C.B. 1983, Ap.J., 272, L1.

# Figure Captions

**Figure 1.** Spectra of the Ly$\alpha$ forest of Q1343+266 A & B, divided by the continuum fit, as a function of vacuum, heliocentric wavelength. The dotted line shows 1-$\sigma$ errors. Tick marks indicate significant absorption features, identified by line numbers which are listed in Table 1 of Paper II. The wavelength of the peak of the quasar Ly$\alpha$ emission line is 3688.0 Å for A and 3689.6 Å for B. The lower panel shows the observed equivalent width threshold (3.5-$\sigma$) for an unresolved line as a function of wavelength, for A (solid line) and B (dashed line).

**Figure 2.** Bayesian probability density, $\mathcal{P}(R)$, is plotted as a function of Ly$\alpha$ cloud radius [solid curve, left-hand vertical scale], in an $(\Omega_o, \Lambda_o) = (1, 0)$ model universe. The integral of this curve over any range in $R$ is the probability that $R$ lies in that range, given the observed pattern of common and uncommon lines ("hits" and "misses") in the spectra of Q1343+266 A & B. Two-sided integrals of $\mathcal{P}(R)$ are also plotted [dashed line], so that one can read off upper and lower bounds to $R$ at any desired level of statistical confidence [right-hand scale]. This plot is for an Einstein-de Sitter cosmology; however constraints on $R$ in other model universes can be found by simply scaling with a numerical factor (see text).

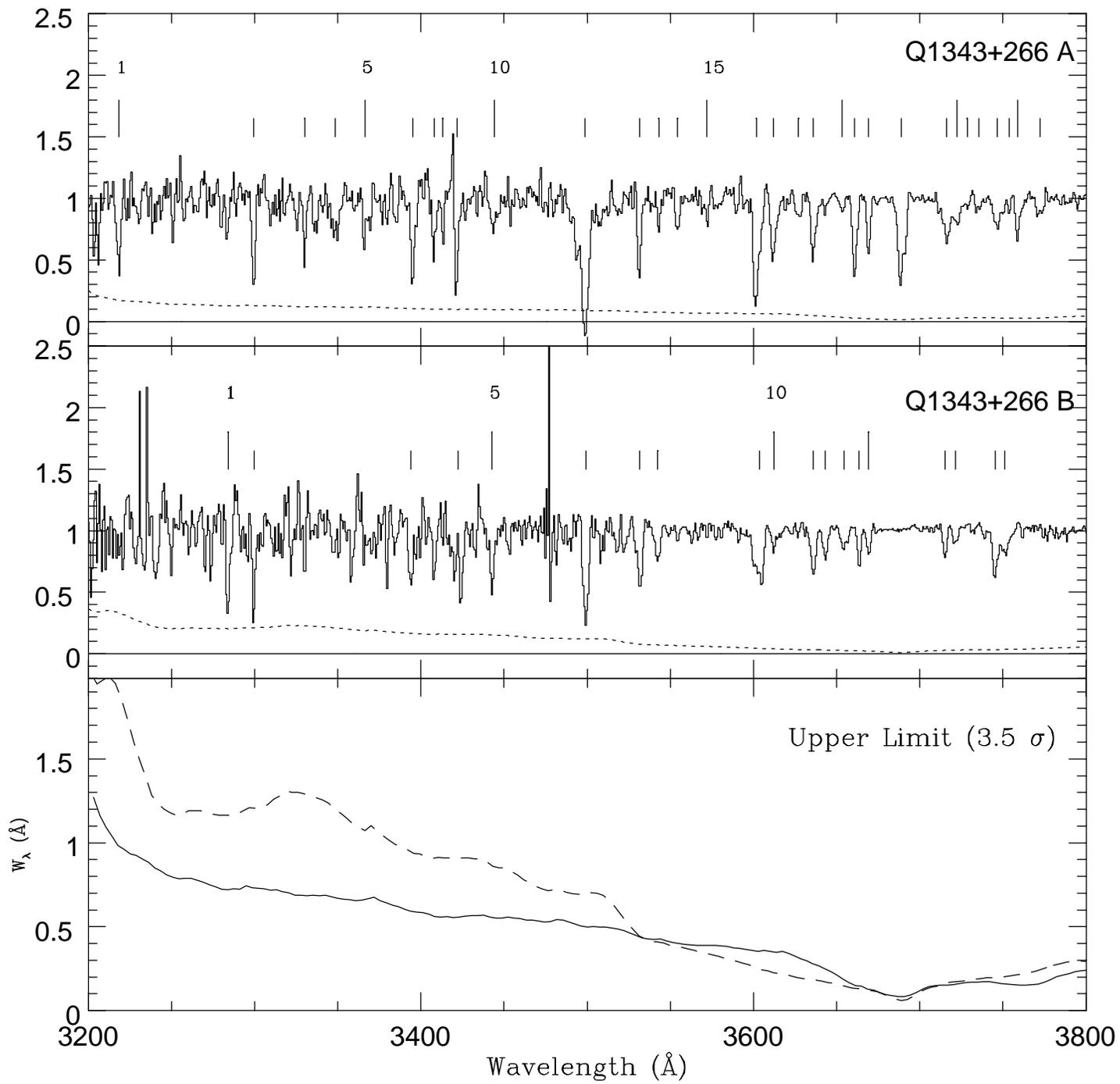

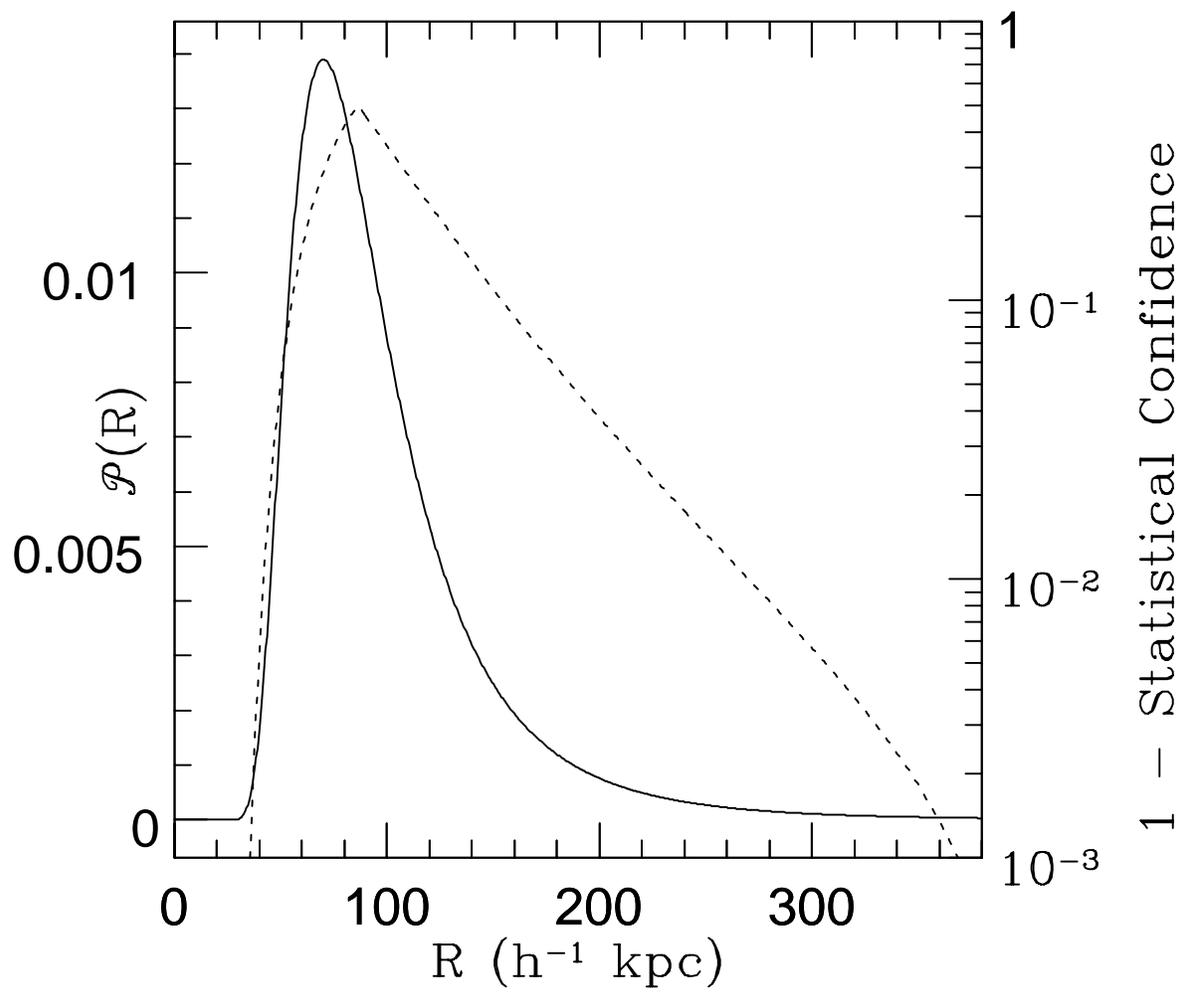

TABLE 1
LINE COINCIDENCES AND SIZE BOUNDS
FOR VARIOUS LY $\alpha$ FOREST SAMPLES

| Sample: $S/N$ Threshold and $\lambda$ Range | No. of Hits | No. of Misses | Cloud Radius Limits ($h^{-1}$ kpc) (99% Confidence) | | |
|---|---|---|---|---|---|
| | | | Lower | Upper | Median |
| 3.5 $\sigma$ | 11 | 4 | 42.5 | 270.0 | 86.5 |
| 3.5 $\sigma$ (no BAL) | 6 | 1 | 40.5 | 377.0 | 115.5 |
| 4.0 $\sigma$ | 9 | 4 | 38.0 | 238.0 | 75.5 |
| 4.0 $\sigma$ (no BAL) | 4 | 1 | 33.5 | 363.0 | 90.5 |
| 4.5 $\sigma$ | 9 | 3 | 40.5 | 308.5 | 89.5 |
| 4.5 $\sigma$ (no BAL) | 4 | 1 | 33.5 | 363.0 | 90.5 |
| 5.0 $\sigma$ | 7 | 3 | 35.5 | 273.0 | 75.5 |
| 5.0 $\sigma$ (no BAL) | 3 | 1 | 29.5 | 350.0 | 76.5 |